\documentclass[pra,twocolumn,showpacs]{revtex4-1}
\usepackage{amsmath,amscd,amssymb,color}
\usepackage{graphicx,amsfonts,epsf}
\usepackage[normalem]{ulem}
\usepackage{epstopdf}
\usepackage{float}
\usepackage{hyperref}
\usepackage{xcolor}

\usepackage{enumerate}
\usepackage{physics}
\usepackage{mathtools}
\begin{document}
\title{Experimental decoherence mitigation 
using a weak measurement-based scheme
and the duality quantum algorithm} 
\author{Gayatri Singh}
\email{ph20015@iisermohali.ac.in}
\affiliation{Department of Physical Sciences, Indian
Institute of Science Education \& 
Research Mohali, Sector 81 SAS Nagar, 
Manauli PO 140306 Punjab India.}
\author{Akshay Gaikwad}
\email{akshay.iiser@gmail.com}
\affiliation{Department of Physical Sciences, Indian
Institute of Science Education \& 
Research Mohali, Sector 81 SAS Nagar, 
Manauli PO 140306 Punjab India.}
\author{Arvind}
\email{arvind@iisermohali.ac.in}
\affiliation{Department of Physical Sciences, Indian
Institute of Science Education \& 
Research Mohali, Sector 81 SAS Nagar, 
Manauli PO 140306 Punjab India.}
\author{Kavita Dorai}
\email{kavita@iisermohali.ac.in}
\affiliation{Department of Physical Sciences, Indian
Institute of Science Education \& 
Research Mohali, Sector 81 SAS Nagar, 
Manauli PO 140306 Punjab India.}
%%%%%%%%%%%%%%%%%%%%%%%%%%%%%%%%%%%%%%%%%%%%%%%%%%%%%%%%%%
\begin{abstract}
We experimentally demonstrate a weak measurement and
measurement reversal-based scheme to ameliorate the effects
of decoherence due to amplitude damping, on an NMR quantum
processor. The weak measurement and measurement reversal
processes require the implementation of non-unitary
operations, which are typically infeasible on conventional
quantum processors, where only unitary quantum operations
are allowed.
The duality quantum algorithm is used to
efficiently implement the required non-unitary quantum
operations corresponding to weak measurement and measurement
reversal.  We experimentally validate the efficacy of the
weak measurement-based decoherence mitigation scheme by
showing state protection on a four-qubit system, with one
qubit being designated as the `system qubit', while the
remaining three qubits serve as `ancilla qubits'. 
Our
experimental results clearly demonstrate the success of the
weak measurement-based decoherence mitigation scheme in
protecting the desired state. Since the measurement
process involved has trace less than unity, the scheme can
be thought of as a filtration scheme, where a subset of
the spins is protected while the rest of the spins can
be discarded.
\end{abstract} 
\maketitle
%%%%%%%%%%%%%%%%%%%%%%%%%%%%%%%%%%%%%%%%%%%%%%%%%%%%%%%%%%%%%%%%
\section{Introduction}
\label{sec1}
A significant hurdle in the physical 
realization of quantum computers is the
deleterious effects of decoherence, which severely hampers their
performance~\cite{Breuer2007}.  Decoherence can lead to substantial errors in
the computational output, leading to diminished experimental fidelity and a
decline in the quality of quantum 
devices~\cite{Rotter-rpp-2015,harper-np-2020}.
Numerous approaches have been suggested to alleviate the impact of decoherence,
such as quantum error correction\cite{knill-pra-1997},
decoherence-free subspaces\cite{lidar-prl-1998,bera-qip-2023},
quantum Zeno effect\cite{wang-prl-2013,facchi-pra-2004,singh-pra-2014}, 
and dynamical decoupling sequences\cite{singh-pra-2017,gautam-ijqi-2023},
all of which have been implemented with varying degrees of success.

A recent innovative approach to protect quantum states against decoherence,
utilizes weak measurements (WM) and measurement reversal (MR)
operations~\cite{koro-pra-2010,wang-pra-2014,wang-pra-2014-2,zou-pra-2017}.
This strategy has proved to be successful in protecting against both amplitude
damping (AD) and generalized amplitude damping (GAD) channels, on optical
systems and on superconducting qubits~\cite{kim-np-2012}.  The efficacy of most
WM-based and MR-based schemes hinges on the reversibility of WM
operations~\cite{koashi-prl-1999,koro-prl-2006}, and has been experimentally
validated using superconducting and photonic
qubits\cite{yong-oe-2009,katz-prl-2008,kim-np-2012}.  Both WM and MR operations
involve non-unitary operators, posing challenges for implementation on
conventional quantum processors. Since both the WM and MR
processes have trace less than unity, the scheme therefore
can be considered to be a filtration process, 
where a subset of the spins undergoing decoherence under the AD
channel are protected against the noise, while the rest of the spins
have to be discarded.

Duality quantum algorithms~\cite{xin-pra-2017} and dilation
algorithms~\cite{gaikwad-pra-2022,kade-prr-2021} are two methods which can
simulate the arbitrary non-unitary dynamics of an open quantum system.  Both
methods rely on a comprehensive understanding of the Kraus operators which
characterize the given quantum channel. The Kraus operators
corresponding to the contractions are non-unitary
operators which preserve or shrink the norm of any
vector~\cite{hu-sc-2020}. The duality quantum algorithm enables the simulation
of non-unitary quantum processes in a single experiment, with the ancilla
system possessing a dimension which is equal to the greater quantity between
the number of Kraus operators and the number of unitary operators into which
these Kraus operators are decomposed \cite{xin-pra-2017}. Dilation techniques, on the other hand,
employ only one ancilla qubit to simulate an arbitrarily dimensional open
quantum system, however, their experimental complexity increases linearly with
the total number of Kraus operators characterizing the quantum channel. The
efficacy of both these simulation methods has been experimentally demonstrated
through the simulation of various non-unitary quantum
processes~\cite{gaikwad-pra-2022,xin-pra-2017}.

In this study, we experimentally demonstrate the efficient use of the duality
quantum algorithm in implementing non-unitary operators corresponding to WM and
MR processes. A generalized 
WM-based and MR-based scheme was used to protect a given quantum
state from decohering under an amplitude damping channel.
The experimental schemes were implemented on a four-qubit NMR quantum
processor and the convex optimization method was used to perform
state tomography. A high fidelity was obtained between the protected
and original states, indicating that the weak measurement-based scheme was
able to successfully protect the state from decoherence under an
amplitude-damping channel.

This paper is structured as follows: In Section~\ref{sec2}, we describe the
scheme to achieve WM-based quantum state protection using the dilation quantum
algorithm. Section~\ref{sec3} contains details of the experimental
implementation of the WM-based state protection scheme on a four-qubit NMR
quantum processor.  Section~\ref{sec4} contains a few concluding remarks.

\section{WM-based state protection scheme using the duality quantum algorithm}
\label{sec2}
\subsection{Action of the amplitude damping channel}
\label{sec2a}
The amplitude damping (AD) channel is a significant noise channel in various
physical systems. In a photonic qubit system, the AD channel arises from photon
loss~\cite{marques-sc-2015}, while in superconducting qubits, it is induced by
zero-temperature energy relaxation~\cite{kim-np-2012}. In NMR systems, the AD
channel is characterized by the spin-lattice relaxation process, also known as
$T_1$ relaxation or longitudinal relaxation~\cite{oliveira}.

Under the AD channel, both diagonal (populations), as well as off-diagonal
(coherences) elements of the density matrix are affected.  Therefore, it is
crucial to develop decoherence mitigation protocols aimed at protecting and
preserving the original quantum state. 

\begin{figure}[h]
\centering
\includegraphics[scale=0.8]{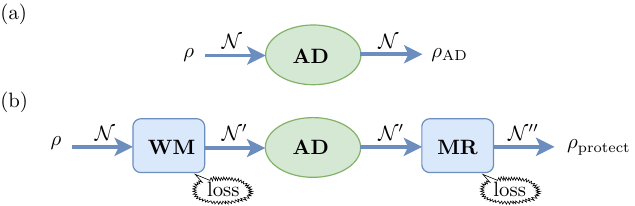}
\caption{Single-qubit state protection strategy using WM and MR: (a) The
quantum state $\rho$ undergoes decoherence under an AD channel.  (b)
The quantum state $\rho$ is protected against decoherence under an AD
channel, by applying WM and MR before and after the AD process,
respectively. The resulting protected state $\rho_{\text{protect}}$ can
be closer to the initial state $\rho$, depending on
the WM and 
MR strengths. Losses occur during the WM and MR operations and
a certain fraction of the spins are lost. $\mathcal{N}$ represents the
total number of spins at the initial step, while $\mathcal{N}'$ and
$\mathcal{N}''$ represent the number of spins after the WM and MR
operations, respectively ($\mathcal{N}''<\mathcal{N}'<\mathcal{N}$).}
\label{fig1}
\end{figure}

Consider the simplest case of a single-qubit system evolving
under the AD channel, where, without loss of generality, the initial state of the
environment is set to be $\ket{0}_E$.  The action of the AD channel on
the system qubit can be characterized by the joint evolution
of the system+environment state as~\cite{kim-np-2012}:

%--------------------------------------
\begin{equation} \label{eq_ad}
\begin{aligned}
& |0\rangle_{\mathrm{S}}|0\rangle_{\mathrm{E}}
\rightarrow|0\rangle_{\mathrm{S}}|0\rangle_{\mathrm{E}} \\
& |1\rangle_{\mathrm{S}}|0\rangle_{\mathrm{E}} \rightarrow
\sqrt{1-p}|1\rangle_{\mathrm{S}}|0\rangle_{\mathrm{E}}+
\sqrt{p}|0\rangle_{\mathrm{S}}|1\rangle_{\mathrm{E}}
\end{aligned}
\end{equation} 
%--------------------------------------
where $0 \leq p \leq 1$ is the strength of the AD channel.
In certain physical scenarios, $p$ can be
expressed as $p = 1-e^{-\gamma t}$ where $\gamma$ is the  
relaxation
rate, typically expressed as $\gamma = 1/{T_1}$. 
It is evident from Eq.(\ref{eq_ad}) that, 
the AD channel affects only the $\ket{1}_S$ component, 
and leaves the
$\ket{0}_S$ component unchanged. 
The system evolution
governed by
Eq.(\ref{eq_ad}), is completely characterized by two Kraus
operators\cite{wang-pra-2014}: 
\begin{equation}
E_0 = \begin{pmatrix} 1 & 0 \\ 0 & \sqrt{1-p}
\end{pmatrix} 
\quad{\rm and} \quad E_1 = \begin{pmatrix} 0 & \sqrt{p} \\ 0 & 0
\end{pmatrix}. 
\end{equation}
Using the Kraus operator decomposition, 
the evolution of the system can be expressed as:
\begin{equation}
\rho_{\text{AD}} =  E_0 \rho(0) E^{\dagger}_0 +  E_1 \rho(0) E^{\dagger}_1
\label{eq1}
\end{equation}
%--------------------------------------
where $\rho(0)=\ket{\Phi}\bra{\Phi}$ is the initial density
matrix (at $t=0$) of the system.  The output density matrix
$\rho_{\text{AD}}$ can be written as:
%-------------------------------------- 
\begin{equation}
\rho_{\text{AD}}=\begin{pmatrix}
p+(1-p) \rho_{11}(0) & \sqrt{1-p} \rho_{12}(0)\\
\sqrt{1-p}\rho_{21}(0) & (1-p) \rho_{22}(0)
\end{pmatrix}
\end{equation}
%--------------------------------------
\subsection{State protection using a WM and MR-based protocol}
\label{sec2b}
The WM+MR protocol for state protection is based on the fact
that the effect of the WM operation can be reversed to a certain
extent by applying the MR operation.
Since both
these processes are non-trace preserving operations, only a
subset of spins remain after these operations are applied,
while the other spins are discarded.
Therefore, these processes correspond to a filter.
Keeping in mind that 
a certain fraction of spins will be discarded, the scheme
shows effective state protection against decoherence by the AD channel,
on the filtered subset of spins.  The system state is
first partially projected onto the $\ket{0}_S$ state using
the WM operator, before subjecting it to the AD channel.
The deleterious effect of the AD channel on the system is
mitigated by the WM operation.  Finally, the  MR operation
reverses the effect of the WM operation, by partially
projecting the system towards the $\ket{1}_S$ state.

The WM-based state protection scheme
is illustrated in Fig.\ref{fig1}.  
Consider an ensemble of  $\mathcal{N}$ spins in the state
$\rho$, evolving under the AD channel, as shown in 
Fig.~\ref{fig1}$\mathrm{(a)}$. 
For the  WM-based
protection scheme depicted in
Fig.~\ref{fig1}$\mathrm{(b)}$, a non-trace-preserving
operator WM is
applied before the system passes through the AD channel and
an MR operation (which is again a  non-trace-preserving
operation) is  applied after the action of the AD channel. 

As a result, spins are lost during these steps, reducing the accessible
ensemble size.  The WM and MR processes are characterized by non-unitary
operators $K_{WM}$ and $K_{MR}$ respectively, given by~\cite{wang-pra-2014}:
%--------------------------------------
\begin{equation} \label{eq4}
\small
K_{WM}= \begin{pmatrix}
1&0\\
0&\sqrt{1-w}
\end{pmatrix} \quad \& \quad K_{MR}= \begin{pmatrix}
\sqrt{1-w_r}&0\\
0&1
\end{pmatrix}
\end{equation}\\
%--------------------------------------
where $w$ and $w_r$ are the strengths of the WM and MR
operations, respectively. After the application of the WM operation, the initial state $\rho$ changes to
\begin{equation}
\begin{split}
        \rho_{\text{wm}}=&\frac{\sigma_{\text{wm}}}{ \text{Tr}[\sigma_{\text{wm}}]}\\
    =&\frac{1}{ \text{Tr}[\sigma_{\text{wm}}]}\begin{pmatrix}
        \rho_{11}(0) &\sqrt{(1-w)}\rho_{12}(0) \\
        \sqrt{(1-w)}\rho_{21}(0) &(1-w)\rho_{22}(0)
        \end{pmatrix}
\end{split}
\end{equation}
where $\text{Tr}[\sigma_{\text{wm}}]=\rho_{11}(0)+(1-w)\rho_{22}(0)$ is the trace of un-normalized density matrix $\sigma_{\text{wm}}$. Following the WM operation, the remaining accessible ensemble size is reduced to $\mathcal{N}' =
\mathcal{N} \, \text{Tr}[\sigma_{\text{wm}}]<\mathcal{N}$, 
with the lost portion treated as spin loss during the WM process. 
The output density matrix after
the entire protection scheme (WM+AD+MR) can be calculated
analytically and is given  by:
%--------------------------------------
\begin{widetext}
\begin{equation}
\rho_{\text{protect}}=\frac{\sigma_{\text{protect}}}{N}=\frac{1}{N}\begin{pmatrix}
(1-w_r)(\rho_{11}(0) + p (1-w) \rho_{22}(0))&
\sqrt{(1-w)(1-p)(1-w_r)} \rho_{12}(0)\\
\sqrt{(1-w)(1-p)(1-w_r)}\rho_{21}(0) & (1-w)(1-p) \rho_{22}(0)
\end{pmatrix}
\label{eq5}
\end{equation}
\end{widetext}
%--------------------------------------
where $N=\text{Tr}[\sigma_{\text{protect}}]$ is a normalization factor and
represents the trace of the un-normalized density matrix
$\sigma_{\text{protect}}$. The MR operation further filters the ensemble,
reducing the protected ensemble size to
$\mathcal{N''}=\mathcal{N}\,\text{Tr}[\sigma_{\text{protect}}]<\mathcal{N'}$
and thus $\mathcal{N-N''}$ spins are lost during the entire process. Therefore
this scheme act as a filter where a subset of spins is filtered out,  whose
states are protected.

Following the application of the protection scheme with prior knowledge of the
damping strength $p$, the optimal value of the measurement reversal strength is
given by $w_r(w,p)=w+p(1-w)=1-e^{-\gamma t}(1-w)$
\cite{koro-pra-2010,kim-np-2012}. The optimal MR strength is calculated by
maximizing the fidelity of the re-normalized final state.  The final state
(Eq.(\ref{eq5})) can be simplified to:
%--------------------------------------
\begin{equation}
\rho_{\text{protect}}=\frac{1}{N}\left[N_1\rho(0)+N_2
\begin{pmatrix}
1&0\\
0&0
\end{pmatrix}\right]
\end{equation}
%--------------------------------------
where
$N=N_1+N_2=(1-p)(1-w)(1+p(1-w)\rho_{22}(0)$, with
$N_1=(1-p)(1-w)$ and $N_2=\rho_{22}(0) (1-w)^2p(1-p)$ also indicates 
the success probability of the protection scheme.  
For a fixed damping strength $p$, the ratio
$N_2/N_1=p(1-w)\rho_{22}(0)$ is a monotonically decreasing
function of $w$. As $w \rightarrow 1$, $N_2/N_1 \rightarrow
0$, indicating that the final state $\rho_{\text{protect}}$
comes closer to the initial state $\rho$. However, this also
implies that the normalization constant $N$ decreases and
approaches 0, indicating a lower success 
probability of protecting the state~\cite{lee-oe-2011}. 

Intuitively, the entire process
(WM-AD-MR) can be thought of as recovering a part of
the entire ensemble and protecting it against decoherence under the
action of the AD channel, with a trade-off
between the fidelity value and 
the success probability. 
Larger
values of WM and MR strengths lead to maximum protection,
however the size of the protected subensemble 
becomes concomitantly smaller.

\begin{figure*}[t]
\centering
\includegraphics[scale=1]{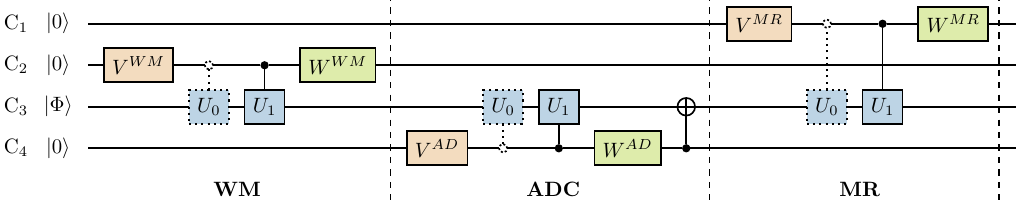}
\caption{Quantum circuit for the WM+MR based state protection scheme. The
	system qubit, denoted by C$_3$, is initialized in the state
	$\ket{\Phi}$, while the ancillary qubits C$_2$, C$_4$, and C$_1$ are
	initialized in the state $\ket{0}$ and are utilized for implementing
	WM, AD, and MR, respectively. The single qubit
	operators $V^i$s and $W^i$s are defined in Eq. \ref{eq12}, while
	$U_0=I$ and  $U_1=Z$  represent 
	identity and the Pauli-$z$ gate respectively.
	}
\label{fig2}
\end{figure*}
%%%%%%%%%%%%%%%%%%%%%%%%%%%%%%%%%%%%%%%%%%%%%%%%%%%%%%%%%%%%%%%%%%
\subsection{Algorithms that mimic non-unitary operations}
\label{sec2c}
From an experimental viewpoint, 
it is nontrivial to implement non-unitary operations 
such as $E_0$, $E_1$, $K_{WM}$ and $K_{MR}$ 
on conventional quantum processors (which only allow
unitary operations). In this subsection, we  describe how the
Sz-Nagy's dilation algorithm (SND) and  the duality quantum algorithm (DQA) can
be used to mimic the action of these non-unitary operators, using
a single ancilla qubit.

\subsubsection*{Sz-Nagy's dilation algorithm}
Generally, any Kraus operator $K_i$, corresponding to 
a given quantum process behaves as a contraction operator \cite{hu-sc-2020},
which either shrinks or preserves the norm of a given vector
$\textbf{\textit{v}}$, i.e.
$\frac{||K_{i}\textbf{\textit{v}}||}{||\textbf{\textit{v}}||} \leq 1$. Hence, one can employ the SND algorithm and construct the corresponding higher-dimensional unitary dilation operator $U^{\rm SND}_{K_i}$
as~\cite{gaikwad-pra-2022}:
%--------------------------------------
\begin{equation}
    U^{\text{SND}}_{K_i}=\begin{pmatrix}
        K_i& \sqrt{I-K_i K^\dagger_i}\\
        \sqrt{I-K^\dagger_i K_i}& -K_i^\dagger
    \end{pmatrix}\\
\end{equation}

Note that, for an $n$-qubit system, $K_i$
is a $2^n \times 2^n$ dimensional operator, and the corresponding unitary
dilation operator $U^{\rm SND}_{K_i}$ is $2^{n+1} \times 2^{n+1}$ dimensional.
So, any $n$-qubit non-unitary process, as long as it is characterized by contraction operators, can be simulated with the help of just a single ancilla qubit.

In our specific scenario, both $K_{WM}$ and $K_{MR}$ 
operators are contraction operators, therefore,  $n$-qubit WM and MR operations can be
separately implemented using only one ancilla qubit. Hence, the protection of
an $n$-qubit state can be achieved in a single experiment using just two
ancilla qubits, using the SND algorithm. However, the experimental
implementation of $U^{\rm SND}_{K_i}$ is not trivial. For instance, in the
single-qubit case, both dilation operators $U^{\rm SND}_{K_{WM}}$ and $U^{\rm
SND}_{K_{MR}}$ will require controlled rotation operations. 

The exact gate
decomposition of these operators is given by: 
%--------------------------------------
\begin{align}
   U^{\rm SND}_{K_{WM}}=&Z_{\pi/2}^1.C_1Y^{21}(2\theta_w)\\
    U^{\rm SND}_{K_{MR}}=&Z_{\pi/2}^1.C_0Y^{21}(2\theta_{w_r})
\end{align}
%--------------------------------------
where $\theta_w$ is defined as $\sin^{-1}{(\sqrt{w})}$, and $\theta_{w_r}$ is
defined as $\sin^{-1}{(\sqrt{w_r})}$. $Z_{\pi/2}^1$ represents a $\pi/2$
rotation along the $z$-axis on qubit 1, and $C_{k}Y^{ij}(\theta)$ represents a
controlled rotation operator, where the $i$ and $j$ superscripts denote the
control and target qubits, respectively, and the $k$ subscript denotes the
state of the control qubit. Further, the implementation of
$C_{k}Y^{ij}(\theta)$, will require two more CNOT gates.

\subsubsection*{The duality quantum algorithm}
The DQA framework enables the simulation of non-unitary
operators in a single experiment. The dimension $d$ of the ancillary system is
determined by the maximum of the following two quantities: the number of Kraus
operators and the number of unitary operators into which the Kraus operators
are expanded. Any non-unitary operator (Kraus operator) can be expanded as a
linear combination of a maximum of $d$ unitary operators. This is known as the
unitary expansion (UE) \cite{zheng-sc-2021}. Mathematically, this can be
written as:$\{K_m\}=\sum_{j=0}^{d-1}\alpha_j U_j$, where $U_j$ are unitary
operators and $\alpha_j$ are complex coefficients.  

The key steps in implementing DQA are as follows:
\begin{itemize} 
\item \emph{Initialization:} The DQA quantum circuit is
initialized with $\ket{\Phi}_s \otimes \ket{0}_a$ where
$\ket{\Phi}_s$ and $\ket{0}_a$ are the state of the main system and
ancillary system, respectively.  
\item \emph{UE parameter assignment:} A unitary
operator $V$ is applied to the ancillary system to create a
superposition state: 
\begin{equation*} \ket{\Phi}_s
\ket{0}_a \rightarrow \ket{\Phi}_s V\ket{0}_a =
\sum_{j=0}^{d-1} V_{j0} \ket{\Phi}_s \ket{j}_a 
\end{equation*}
The elements of the first column of the unitary matrix $V$ ($V_{j0}$)
are determined using UE parameters $\alpha_j$, and rest of the
column is obtained using Gram–Schmidt orthogonalization.  
\item \emph{UE terms generation:} Unitary operations $U_j$s are
performed on the system qubit with the state of the ancilla
qubits being the control, with the state evolution being given by:
\begin{equation}
\hspace{1cm} \sum_{j=0}^{d-1} V_{j0} \ket{\Phi}_s
\ket{j}_a \rightarrow \sum_{j=0}^{d-1} V_{j0} (U_j
\otimes \ket{j}_a \bra{j}_a)  \ket{\Phi}_s \ket{j}_a
\label{eq_ue} 
\end{equation} 
These unitary operations $U_j$'s
correspond to the decomposition of $K_m$ and the total effect of these
controlled operators is to generate the UE-terms.  
\item \emph{Superposition:}
A unitary operator $W$ is then applied to the ancillary system to
achieve the final superposition, resulting in:
\begin{equation*} 
 \hspace{1cm}\sum_{j=0}^{d-1} V_{j0}
 U_j\ket{\Phi}_s \ket{j}_a\rightarrow \sum_{j,m=0}^{d-1}
 W_{mj}V_{j0} U_j\ket{\Phi}_s \ket{m}_a \label{eq10}
 \end{equation*} 
The elements of the matrix $W$ are uniquely
determined by using the matrix $V$, such that 
the evolution of non unitary
operator $\{K_m\}$ satisfies the relation $K_{m}=\sum_{j=0}^{d-1} W_{m j} V_{j
 0} U_{j}$ 
\item \emph{Measurement:} With the ancillary system being in the state
$\ket{m}\bra{m}$, measurement on the system qubit will result in $K_m
\ket{\Phi}\bra{\Phi}K_m^\dagger$ and the desired action is simulated.
\end{itemize}   
%-------------------------------------- --------------------------------------

The UE of Kraus operators corresponding to the WM, MR and AD 
channels is given by:
\begin{equation}
\begin{split}
\hspace{1cm}\text{WM}&: \hspace{0.1cm} K_{WM}=a_1^2 I+ b_1^2 Z \\
\hspace{1cm}\text{MR}&: \hspace{0.1cm}  K_{MR}=a_2^2 I- b_2^2 Z\\
\hspace{1cm}\text{AD}&: \hspace{0.1cm}  E_0=a_3^2 I+ b_3^2 Z
\hspace{0.15cm} {\rm and}  \hspace{0.15cm}
E_1=\frac{\sqrt{p}}{2}\,X( I- Z)
\end{split}\label{eq11}
\end{equation}

%--------------------------------------
From Eq.(\ref{eq11}), it is evident that 
the Kraus operators corresponding to both the WM
and MR processes have two UE terms, requiring one ancilla for each process.
The
AD channel on the other hand, has two Kraus operators, 
each having two UE terms and also requires only
one ancilla. From Eq.(\ref{eq11}), the unitary operators
$U_j$s in Eq.(\ref{eq_ue})
can be set to $U_0=I$ and $U_1=Z$, to simulate 
the action of these non-unitary operation. For the 
Kraus operator of WM ($K_m = K_{WM} $), MR ($K_m = K_{MR} $) and 
the AD channel ($E_0$ and $E_1$), 
the corresponding $V^{i}$ as well as  $W^{i}$ are given as follows,
%--------------------------------------
\begin{eqnarray}
V^{i}=\begin{pmatrix}
a_i&-b_i^*\\
b_i&a_i
\end{pmatrix} \quad \text{and} \quad W^{i}=\begin{pmatrix}
a_i&b_i^*\nonumber \\
c_i/a_i&-c_i/b_i
\end{pmatrix}, \\
a_i = \sqrt{\frac{1 + \sqrt{1 -\alpha_i}}{2}};\, b_i = \pm \sqrt{\frac{1 - \sqrt{1 -\alpha_i}}{2}};
\, c_i=\frac{\sqrt{\alpha_i}}{2}\label{eq12}
\nonumber \\
\end{eqnarray} 
%--------------------------------------
where $i=WM,MR$ and $AD$, with $\alpha_{WM}=w$, $\alpha_{MR}=w_r$ and
$\alpha_{AD}=p$. For the WM operation and the AD channel, the sign of $b_i$ is
positive, while for the MR operation, the sign of $b_i$ is negative.

When implementing the SND algorithm, a higher dimensional 
unitary matrix is required to implemented on 
the combined system-ancilla state, 
in order to simulate the action of one Kraus operator. 
The action of the given Kraus operator is 
then simulated in a subspace spanned by a few computational basis states. 
For the AD channel, the simultaneous action of the
corresponding Kraus operators $E_0$ and $E_1$ is simulated
in the higher-dimensional space spanned
by (ancilla qubit+ system qubit) $\{ \ket{0}_a  \ket{0}_s,
\ket{0}_a  \ket{1}_s,  \ket{1}_a  \ket{0}_s,
\ket{1}_a  \ket{1}_s  \}$ :
\begin{equation}
\ket{0}_a\ket{\Phi}_s \xrightarrow{\text{AD}}  \ket{0}_a E_0\ket{\Phi}_s+
\ket{1}_a E_1\ket{\Phi}_s
\end{equation}
Hence, measurement on the system qubit is sufficient 
to yield the evolution result of AD channel ~\cite{wei-scpma-2018}. 
For the WM and MR operators, we only need to simulate a single Kraus
operator. The action of corresponding Kraus operators, $K_{WM}$ and $K_{MR}$,  is simulated in the higher-dimensional subspace
(ancilla qubit+ system qubit) spanned by $\{ \ket{0}_a
\ket{0}_s, \ket{0}_a  \ket{1}_s  \}$:
\begin{equation}
\begin{split}
\ket{0}_a\ket{\Phi}_s \xrightarrow{\text{WM}} &\ket{0}_a
K_{WM}\ket{\Phi}_s+ \ket{1}_a (a_1^2 I-b_1^2 Z)\ket{\Phi}_s\\
\ket{0}_a\ket{\Phi}_s \xrightarrow{\text{MR}} &\ket{0}_a
K_{MR}\ket{\Phi}_s+ \ket{1}_a (a_2^2 I+b_2^2 Z)\ket{\Phi}_s\\
\end{split}\label{eq14}
\end{equation} 
Therefore, the simulated output density matrix 
can be recovered by measuring only those elements 
which spans the subspace. 

It turns out that the action of the non-unitary operators $K_{WM}$ and $K_{MR}$
can each be implemented using only one CNOT gate in the single-qubit case,
using the DQA.  This is in contrast to the SND algorithm, making the DQA
experimentally less resource-intensive and more efficient. Due to this
advantage, we opted to experimentally implement non-unitary processes using the
DQA and have demonstrated its application in the quantum state protection
scheme. We note here in passing that the DQA circuit presented in
Fig.~\ref{fig2} can also be applied to other WM and MR-based schemes.

%%%%%%%%%%%%%%%%%%%%%%%%%%%%%%%%%%%%%%%%%%%%%%%%%%%%%%%%%%%%%%%%%%%%%%
\section{Experimental implementation on an NMR quantum processor}
\label{sec3}
\subsection{Realizing NMR qubits}
\label{sec3a}
%--------------------------------
\begin{figure}
\centering
\includegraphics[scale=0.9]{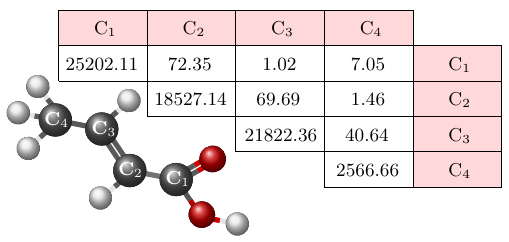}
\caption{The molecular structure and Hamiltonian parameters:
chemical shift (in ppm) and scalar J-coupling strengths (in
Hz), are tabulated for the four carbon atoms labeled as
C$_1$,C$_2$,C$_3$ and C$_4$ in $^{13}$C-labeled
trans-crotonic acid. Within the table, the rows display the
chemical shift in the diagonal entries, while the
off-diagonal entries indicate the scalar J-coupling between
respective nuclei. In the diagram of the molecule, the red
colored spheres represent oxygen nuclei, while the white
spheres represent hydrogen nuclei.}
\label{fig3}
\end{figure}
%--------------------------------------
The four $^{13}$C nuclei of $^{13}$C-labeled trans-crotonic
acid,  dissolved in acetone-D6, were used to realize a
four-qubit system. The schematic of the molecule, along with
NMR Hamiltonian parameters such as chemical shift $\nu_i$
(in ppm) and scalar J-coupling $J_{i j}$ (in Hz) are
depicted in Fig.~\ref{fig3}.  During the experiment, a
broadband decoupling sequence, WALTZ-16
\cite{shaka-jmr-1983}, was applied to decouple the methyl
group and other proton nuclei. All experiments were
performed on a Bruker Avance-III 600 MHz FT-NMR
spectrometer, equipped with a standard 5mm QXI probe, at
room temperature ($\sim 300K$).

In the rotating frame, the NMR Hamiltonian of four spin 1/2
nuclei under the weak coupling approximation can be
expressed as~\cite{oliveira}:
%--------------------------------------
\begin{equation}
\mathcal{H}=-\sum_{i=1}^{4} 
(\omega_{i}-\omega_{{\rm rf}}) I_{i z}+
\sum_{i<j,j=1}^{4} 2 \pi J_{i j} 
I_{i z} I_{j z}
\label{eq13}
\end{equation}
%--------------------------------------
where $\omega_{\text{rf}}$ is the frequency of the rotating
frame.  The spins are denoted by the index $i$, where
J$_{ij}$ represents the scalar coupling between the $i$th
and $j$th spins. Additionally, $\omega_i=2\pi\nu_i$ and
$I_{iz}$ represent the Larmor frequency and the
$z$-component of the spin angular momentum of the $i$th
spin, respectively.  More details of the molecular
parameters and the T$_1$ and T$_2$ relaxation rates can be
found in the Ref.~\cite{gayatri-qip-2023}.
%--------------------------------
\begin{figure}[h!]
\centering
\includegraphics[scale=1.15]{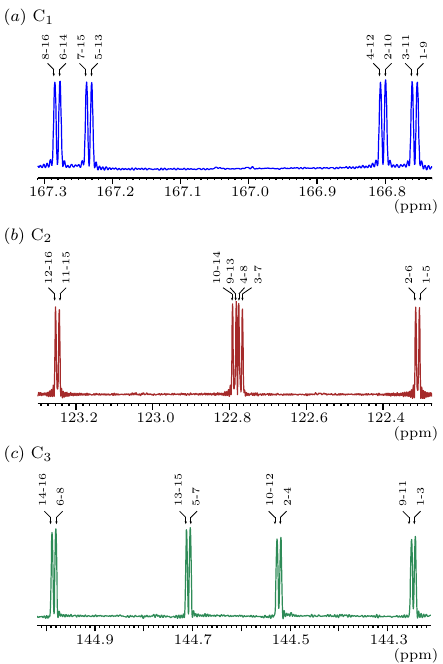}
\caption{NMR spectrum of qubits C$_1$,C$_2$ and C$_3$ obtained after
implementing a $\pi/2$ pulse on the thermal equilibrium state. The spectral
lines of each qubit are labelled by the corresponding
transition in the density matrix.}
\label{fig4}
\end{figure}
%--------------------------------------
\subsection{Experimental implementation of the state protection scheme}
\label{sec3b}
To experimentally demonstrate the efficacy of the WM-based state protection
scheme using DQA, we have performed two sets of experiments. In one set, we
only simulate the action of the AD channel using DQA with
the help of one ancilla qubit and in the second set of
experiments we perform state protection, i.e., apply the WM
and MR operations before and after the AD channel.  In our
experiments, we labeled C$_3$ as the system qubit and C$_2$,
C$_4$, and C$_1$ as ancillary qubits for the WM, AD and MR
processes, respectively.  We prepare the
four-qubit system in the state
$\ket{0}\otimes\ket{0}\otimes\ket{\Phi}\otimes\ket{0}$ and
implement the quantum circuit depicted in Fig.~\ref{fig2}.
The density operator of system qubit is
represented by $\rho=\ket{\Phi}\bra{\Phi}$.

The quantum circuit for simulation of the AD channel via DQA is given in the
second block of Fig.~\ref{fig2}. We have used GRAPE (Gradient Ascent Pulse
Engineering) optimized rf-pulses \cite{khaneja-jmr-2005,tosner-jmr-2009} to
implement the quantum circuit and for the AD channel, the optimized pulse
length was $\approx 25$ ms. 

Measurement only on the system qubit is sufficient to 
capture the evolution of the entire quantum channel. 
We reconstructed the density matrix
$\rho_{AD}$ using constrained convex
optimization (CCO) based quantum state tomography (QST)
method\cite{gaikwad-qip-2021,gaikwad-qip-2022}, using two
tomographic operations $\{IIII, IIXI\}$, where $I$ is the
identity operator and $X$ denotes a spin-selective $\pi/2$
rotation along the $x$-axis. The unitary operators
corresponding to these tomographic operations were optimized
using the GRAPE technique. The average fidelity of all
single-qubit GRAPE pulses is $\geq 0.997$, and they are
crafted to be robust against rf inhomogeneity, with a
duration ranging from $350$ to $370\mu$~s. The state
fidelity was calculated using the Uhlmann-Jozsa fidelity
measure~\cite{uhlmann-rpmp-1976,jozsa-jmo-1994}.

During the implementation of  the state protection scheme,
the system qubit is initially subjected to a weak
measurement, followed by decoherence under the amplitude
damping channel.  Subsequent to undergoing decoherence, a
measurement reversal operation is applied to the system
qubit.  

The quantum circuit in
Fig.~\ref{fig2} was implemented on four NMR qubits and GRAPE-optimized rf
pulses were employed to implement the unitaries constructed using DQA for the
Kraus operators corresponding to the WM, MR and AD channel. The implementation
of the WM, MR operations and the AD channel required approximately 8 ms, 27 ms
and 25 ms, respectively. Consequently, the overall pulse length, including
pseudo-pure state (PPS) preparation (the NMR pulse sequence for PPS
preparation, can be found in the reference~\cite{gayatri-qip-2023}), was
approximately 102 ms. The experiments were performed with varying time values
$t$, with fixed weak measurement strength $w$ and damping strength $\gamma$,
for different input states namely,
$\ket{\Phi_1}=\frac{\sqrt{3}}{2}\ket{0}+\frac{\iota}{2}\ket{1}$,
$\ket{\Phi_2}=\ket{-}=\frac{1}{\sqrt{2}}\ket{0}+\frac{\iota}{\sqrt{2}}\ket{1}$ and
$\ket{\Phi_3}=\ket{1}$. The experiments were then repeated by varying the weak
measurement strength $w$,  keeping $\gamma$ and $t$ unchanged. The optimal
measurement reversal strength was chosen to be $w_r=p+w(1-p)$ for given $(p,w)$ value.

The protected state of the system qubit can be recovered from the four qubit
subspace spanned by:~$\ket{0000},\ket{0001},\ket{0010}$ and $\ket{0011}$.  The
reconstruction of the final density matrix $\rho_{\rm protect}$ 
is obtained by
reducing the 4-qubit $16\times 16$ density matrix $\sigma$, with respect to
qubit~4(C$_4$) and then projecting it onto the smaller subspace spanned by $\{
\ket{000},\ket{001} \}$, which is equivalent to estimating a $2\times 2$
partial density matrix (corresponding to the first two rows and columns of the
reduced density matrix $\text{Tr}_{\text{C}_4}[\sigma]$ ) given by:

%--------------------------------------
\begin{equation} \label{output}
    \rho_{\rm protect}= \frac{1}{N}\begin{pmatrix}
        \sigma_{11}+ \sigma_{22} &   \sigma_{13}+ \sigma_{24}\\
         \sigma_{13}^*+ \sigma_{24}^* &  \sigma_{33}+ \sigma_{44}
    \end{pmatrix}
\end{equation}
%--------------------------------------
where $N=\sigma_{11}+ \sigma_{22}+\sigma_{33}+ \sigma_{44}$ is the
normalization constant and $\sigma_{ij}$s are the elements of 
the four-qubit density
matrix $\sigma$, obtained after implementing the quantum circuit given in
Fig.~\ref{fig2}.

%--------------------------------
\begin{figure*}[ht!] 
\centering 
\includegraphics[scale=1]{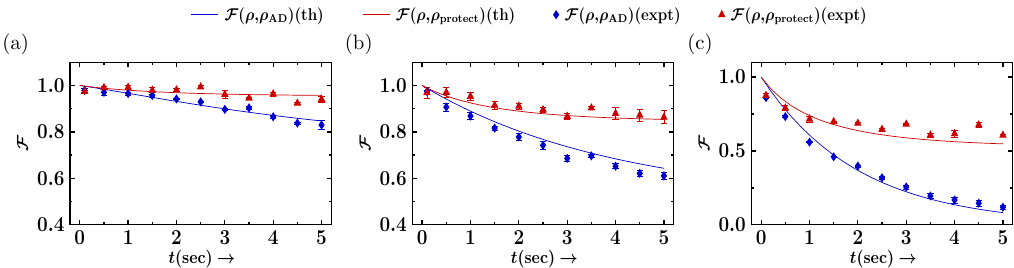}
\caption{The plots in panels (a), (b), (c) compare the theoretical (solid
curve) and experimentally measured (points with error bars) fidelity of
the density matrices $\rho_{\text{AD}}$ and $\rho_{\text{protect}}$,
respectively for the input state $\ket{\Phi_1}$, $\ket{\Phi_2}$ and
$\ket{\Phi_3}$, respectively , over different times $t$. In these
experiments, the WM strength was set to $w=0.1$ and the damping
strength to $\gamma=0.5$. The blue data points correspond to the
scenario where the system qubit undergoes decoherence solely under the
amplitude damping (AD) channel, while the red data points represent the
results obtained after implementing the state protection scheme using
the WM and MR operations.}
\label{fig5} 
\end{figure*}
%--------------------------------------

Since the action of WM and MR operation are
simulated in $\ket{0}$ subspace only, the measurement on
system qubit will not yield the necessary information. The
full QST of a four-qubit density matrix  $\sigma$ requires a
set of 15 tomographic pulses. However it turns out that, the
density matrix $\rho_{{\rm protect}}$
characterizing the state of the system qubit can be
reconstructed using a set of only four tomographic pulses:
$IIII,YIII,IYII,IIYI$.  The off-diagonal entries,
$\sigma_{13}$ ($\sigma_{13}^*$) and $\sigma_{24}$
($\sigma_{24}^*$) in Eq.(\ref{output}) can be directly
determined by measuring the sum of  line
intensities of the rightmost peak and its neighbor  of the
C$_3$ qubit~(Fig. \ref{fig4}$\mathrm{(c)}$). The real and
imaginary parts of these readout elements are proportional
to the line intensity of the absorption and dispersion mode
peaks, respectively.

The diagonal entries of the density matrix $\rho_{\rm
protect}$ in Eq.(\ref{output}) can be directly computed by
applying the set of tomographic pulses $YIII,IYII,IIYI$,
followed by signal acquisition on C$_1$, C$_2$ and C$_3$
qubits as follows:

%--------------------------------------
\begin{equation}
\begin{split}
        \sigma_1=YIII.\sigma.YIII^\dagger\\
         \sigma_2=IYII.\sigma.IYII^\dagger\\
          \sigma_3=IIYI.\sigma.IIYI^\dagger\\
\end{split}
\end{equation}
%--------------------------------------
The NMR spectra of the qubits C$_1$, C$_2$ and C$_3$  
have 8 peaks each (Fig. \ref{fig4}). The diagonal 
entries are evaluated by:
%--------------------------------------
\begin{equation}
    \begin{split}
    \sigma_{11}+ \sigma_{22}=& (1+2 \alpha+ 4\beta +8 \gamma)/8\\
    \sigma_{33}+ \sigma_{44}=& (1+2 \alpha+ 4\beta -8 \gamma)/8
    \end{split} \label{eq20}
\end{equation}
%--------------------------------------
Here, $\alpha$ denotes the sum of line intensities (in
absorption mode) corresponding to the state $\sigma_1$
across all eight peaks of qubit C$_1$.  Similarly, $\beta$
represents the sum of line intensities (in absorption mode)
for the state $\sigma_2$ across the rightmost four peaks of
qubit C$_2$, and $\gamma$ represents the sum of line
intensities (in absorption mode) of first rightmost peaks
and the second neighboring peak to the rightmost peak across
qubit C$_3$. It is important to note that since all the
elements are computed independently, the final
reconstructed density matrix may not represent a valid physical state.
Here, we have used constrained convex optimization (CCO) QST
method to ensure the reconstruction of a valid density matrix
\cite{gaikwad-qip-2021}.

In Fig.\ref{fig5}, the panels $\mathrm{(a)}$,$\mathrm{(b)}$ and
$\mathrm{(c)}$ depict a comparison between the fidelity of
density matrices $\rho_{\text{AD}}$ and
$\rho_{\text{protect}}$ at $w=0.1$ and
$\gamma=0.5$ with respect to the input state
$\ket{\Phi_1},\ket{\Phi_2}$ and $\ket{\Phi_3}$,
respectively. Theoretical (solid curve) as well as
experimental results (points with error bar) are shown over
a time range from $t=0.1$ to $t=5$~s in
both scenarios: when the system qubit undergoes the AD
channel solely (blue) and when the system qubit undergoes
the protection scheme using WM and MR (red).
We did not consider the limiting cases when
$\alpha_i=0$ or $1$ due to the choice of $W^i$ (Eq.
\ref{eq12}), as the denominator tends to zero. The system
qubit was prepared in state $\ket{\Phi_1},\ket{\Phi_2}$ and
$\ket{\Phi_3}$ with fidelity 
$0.9895\pm0.0033$, $0.9914\pm0.0029$ and  $0.9955   \pm0.0015$ respectively.
For instance, the density matrix of the system qubit
initialized in state $\ket{\Phi_2}$ 
is given as,
\begin{equation}
\small
   \rho(0)=
   \begin{pmatrix}
       0.4411\pm 0.0230 & \begin{array}{c}
           -0.0680\pm 0.0114\\
            -(0.4914 \pm 0.0029)\iota 
       \end{array}\\
       \begin{array}{c}
           -0.0680\pm 0.0114\\
            +(0.4914 \pm 0.0029)\iota \end{array}& \small{0.5589\pm0.0230}
    \end{pmatrix}
\end{equation}
At  $t=5$~s, the state fidelities of the system qubit undergoing
the AD channel were  $0.8296\pm0.0193$, $0.6106\pm0.0156$
and $0.1189\pm0.0153$, while the fidelities after
implementing the protection scheme are
$0.9402\pm0.0123$,  $0.8631\pm0.0279$ and $0.6058\pm0.0052$ for the states
$\ket{\Phi_1},\ket{\Phi_2}$ and $\ket{\Phi_3}$, respectively.
For the input state $\ket{\Phi_2}$ undergoing the AD
channel and the protection scheme, the density matrices at time
$t=5$~s are:           \begin{equation} \small \begin{split}
\rho_{AD}&= \begin{pmatrix} 0.9402\pm0.0107 &
\begin{array}{c}0.0107\pm0.0113 \\-(0.1106\pm0.0156)\iota
\end{array}\\ \begin{array}{c}0.0107\pm0.0113
\\+(0.1106\pm0.0156)\iota \end{array}& 0.0598\pm0.0107
\end{pmatrix}\\ \rho_{protect}&= \begin{pmatrix} 0.6671\pm
0.0539 & \begin{array}{c}-0.0941\pm0.0437 \\-(0.3631\pm
0.0280)\iota \end{array}\\ \begin{array}{c}-0.0941\pm0.0437
\\+(0.3631\pm 0.0280)\iota \end{array}& 0.3330\pm 0.0539
\end{pmatrix} \end{split}\label{eq23} \end{equation}
From Eq.~\ref{eq23}, one can observe that
as $t$ increases, the damped state becomes closer to
the $\ket{0}$ state. This implies that the states which have
a higher probability of being in the state $\ket{1}$ 
get damped faster than those close to the state
$\ket{0}$. For instance, the state
$\ket{\Phi_2}$ has equal probability of being in state
$\ket{0}$ and $\ket{1}$, and exhibits a slower damping rate
(\ref{fig5}$\mathcal(b)$) 
	than the state $\ket{\Phi_3}$ which has
only the $\ket{1}$ component and damps very quickly, causing the
fidelity to tend towards zero (\ref{fig5}$\mathcal(c)$). 
In
contrast, the state $\ket{\Phi_1}$, which has a higher
probability of being in $\ket{0}$  than $\ket{1}$ has
the slowest damping rate amongst the three states
(\ref{fig5}$\mathcal(a)$).

\begin{figure}[t]
\centering
\includegraphics[scale=1]{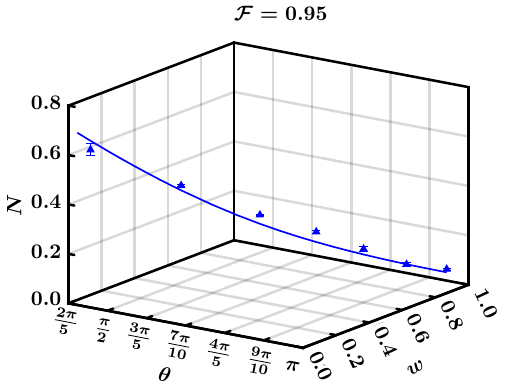}
\caption{A 3D plot of the trace of the output density matrix ($N$) as a
function of WM strength $w$ with AD channel strength  fixed at $p\sim
0.4$ ($\gamma=0.5$, $t=1$), aimed at achieving a maximum fidelity of
0.95 for the input state $\ket{\Phi}=\cos
\frac{\theta}{2}\ket{0}+\iota\sin \frac{\theta}{2}\ket{1}$. The solid
curve represents the theoretical results, while the triangles with
error bars represent the experimental results.  }
\label{fig6}
\end{figure}

We examined the trace of the output density matrix ($N$)
for a general single-qubit state of the form
$\ket{\Phi}=\cos{\frac{\theta}{2}}\ket{0}+\iota\sin{\frac{\theta}{2}}\ket{1}$
across various values of $\theta$, while varying WM strength $w$ (Fig.
\ref{fig6}). Our goal was to achieve a fidelity
$\mathcal{F}=0.95$ of the protected state with respect to initial state,
with a constant AD channel strength $p\sim 0.4$ ($\gamma=0.5$, $t=1$).
We explored values of $\theta$ in the range $0.4225 \,\pi\leq \theta <
\pi$, since for $\theta<0.4225 \,\pi$, fidelity greater than 0.95 is
already achieved with success probability greater than $\approx$ 0.69
even when WM strength is 0. However, as $\theta$ approaches $\pi$, the
required WM strength also increases, which leads to a corresponding
decrease in success probability. 
This plot illustrates the trade-off between
the success probability and the protected state fidelity in
a realistic implementation of the scheme. 
This analysis provides
insights into the effectiveness of the state protection scheme for
different initial states and measurement strengths.
Figure~\ref{fig6} provides basic data about
how the filtered ensemble size depends upon the choice of the state
on the Bloch sphere, keeping the desired fidelity to be fixed at
a minimum of
0$.95$. For an unknown quantum state, one would have to
make a suitable choice, based on the average behavior of
the states on the Bloch sphere, after deciding on a
certain value as a cut off for the fidelity.

The experimental results are in good agreement with the theoretical
simulations, which clearly demonstrates the successful implementation
of the protection scheme with a high fidelity upto time $t=5$~s. As we
increase $t$ or $w$ further, the trace of un-normalized density matrix
approaches to zero, implying that the elements of  the un-normalized
density matrix are getting closer to zero.  In NMR, achieving precision
beyond the first decimal place is challenging; hence, minor deviations
between experimental and theoretical values can lead to significant
changes as the normalization constant (which is significantly smaller
than one). Despite these experimental constraints, we achieved a good
agreement between the theoretical and experimental results by measuring
only the sub-space elements instead of performing full state
tomography. 

We note here that physically separating the protected sub-ensemble is not
possible in our experiments.  Nevertheless, we can still utilize the protected
sub-ensemble of spins for further applications by implementing the desired
operations on the system qubit after the state protection scheme (continuation
of quantum circuit given in Fig.\ref{fig2}). During readout, although the final
NMR signal will originate from all spins (both protected and unprotected), the
fractional contribution from only the protected sub-ensemble can be acquired
using Eqs.~\ref{output} and \ref{eq20}. This approach effectively confines us
to a new computational subspace spanned by $\ket{0000}$, $\ket{0001}$,
$\ket{0010}$, and $\ket{0011}$, where the protected sub-ensemble lives.  
%%%%%%%%%%%%%%%%%%%%%%%%%%%%%%%%%%%%%%%%%%%%%%%%%%%%%%%%%%%%%%%%%%%%%%
%--------------------------------
\section{Concluding Remarks}
\label{sec4}
We experimentally demonstrated a scheme for quantum state
protection, based on weak measurements and measurement
reversal, on a four-qubit NMR quantum information processor.
The experimental implementation included the simulation of
non-unitary processes such as weak measurements, amplitude
damping, and measurement reversal, using the duality quantum
algorithm.  We also provided a generalized quantum circuit
which can be implemented on other quantum hardware.
Experiments were conducted with various
input states under two scenarios: one where the system
solely undergoes the AD channel and another where the
protection scheme is applied after the action of the
AD channel. We used the convex
optimization method to perform tomography of the
protected state and compared it with the original state.
Comparisons between experimental and
theoretical results were made for different cases: keeping
WM strength $w$ and AD strength $p$ constant while varying
time $t$ and keeping time $t$ and AD strength $p$ constant
while varying WM strength $w$. The high experimental
fidelity obtained between the protected state and the
original state clearly demonstrated the successful
implementation of the weak measurement-based quantum state
protection scheme using the duality quantum algorithm.
We also highlighted the trade-off that
while the protected state can become closer to the initial
state, the success probability of doing so decreases
significantly. The scheme is primarily applicable to the
amplitude damping and the generalized amplitude damping
channels, and the efficacy of this scheme against other
noisy processes requires further exploration.

%%%%%%%%%%%%%%%%%%%%%%%%%%%%%%%%%%%%%%%%%%%%%%%%%%%%%%%%%%%%%%%%%%%%%%
\begin{acknowledgements}
All experiments were performed on a Bruker Avance-III 600 MHz FT-NMR
spectrometer at the NMR Research Facility at IISER Mohali.  G.S.
acknowledges University Grants Commission (UGC), India, for financial
support. 
\end{acknowledgements}

%\bibliographystyle{apsrev4-1}
%\bibliography{sp}

\end{document}